\documentclass[usegraphicx,useAMS,usenatbib]{mn2e}
\usepackage{url}

\title[Ram pressure stripping with turbulent ISM]
{Ram pressure stripping in elliptical galaxies: I. the impact of the 
interstellar medium turbulence}
\author[M.-S. Shin \& M. Ruszkowski]
  {Min-Su~Shin,$^1$\thanks{E-mail: 
  msshin@umich.edu, 
  mateuszr@umich.edu}
  Mateusz~Ruszkowski$^{1,2}$\\
  $^1$Department of Astronomy, The University of Michigan, 500 Church Street, Ann Arbor, MI 48109, USA\\
  $^2$The Michigan Center for Theoretical Physics, 3444 Randall Lab, 450 Church St, Ann Arbor, MI 48109, USA
  }
\date{Released 2009 Xxxxx XX}

\begin{document}

\date{Accepted ... Received ..; in original form ..}

\maketitle

\begin{abstract}
Elliptical galaxies contain X-ray emitting gas that is subject to continuous ram pressure stripping over timescales comparable to cluster ages.
The gas in these galaxies is not in perfect hydrostatic equilibrium. Supernova feedback, stellar winds, or active galactic nuclei (AGN) feedback can 
significantly perturb the interstellar medium (ISM). Using hydrodynamical simulations, 
we investigate the effect of subsonic turbulence in the hot ISM on the ram pressure stripping process in early-type galaxies.
We find that galaxies with more turbulent ISM produce longer, wider, and more smoothly distributed tails of the stripped ISM 
than those characterised by weaker ISM turbulence. 
Our main conclusion is that even very weak internal turbulence, 
at the level of $\la 15\%$ of the average ISM sound speed, can significantly 
accelerate the gas removal from galaxies via ram pressure stripping. 
The magnitude of this effect increases sharply with the strength of turbulence. 
As most of the gas stripping takes place near the boundary between the ISM and the intracluster medium (ICM),
the boost in the ISM stripping rate is due to the ``random walk'' of the ISM from the central regions of the galactic potential well
to larger distances, where the ram pressure is able to permanently remove the gas from galaxies. 
The ICM can be temporarily trapped inside the galactic potential well due to the mixing of the turbulent ISM with the ICM. The galaxies with 
more turbulent ISM, yet still characterised by very weak turbulence, can hold larger amounts of the ICM.
We find that the total gas mass held in galaxies decreases with time slower than the mass of the original ISM, 
and thus the properties of gas retained inside galaxies, such as metallicity, can be altered by the ICM over time. 
This effect increases with the strength of the turbulence, and 
is most significant in the outer regions of galaxies.
\end{abstract}

\begin{keywords}
hydrodynamics –- methods: numerical –- galaxies: clusters: general –-
galaxies: evolution –- galaxies: ISM –- galaxies: intergalactic medium
\end{keywords}

\section{INTRODUCTION}

Ram pressure stripping removes gas from galaxies moving relative to the ICM \citep{1972ApJ...176....1G}.
Numerous theoretical studies examined the consequences of this effect for the galaxy and cluster evolution 
by quantifying the amount of gas loss from galaxies, including star formation in galaxies and their ram-pressure 
stripping tails, and determining the metal enrichment of the ICM by stripping 
metal-rich gas from galaxies \citep[e.g.,][]{1994ApJ...437...83B,
2006A&A...452..795D, 2006MNRAS.369.1021M,hester06,2007MNRAS.380.1399R,2007ApJ...671.1434T,
2008MNRAS.383..593M,2008MNRAS.388L..89R,2008ApJ...684L...9T,2008MNRAS.389.1405K,
2009A&A...499...87K,2009A&A...500..693J,2010ApJ...716..810B,2010MNRAS.408.2008T,2011ApJ...729...11K,2011MNRAS.415..257N}.

Previous theoretical investigations of ram pressure stripping did not include all non-thermal energy components in the ISM and ICM. 
In general, non-thermal components include turbulent kinetic energy, magnetic fields, and cosmic-rays. 
Only few theoretical studies \citep[e.g.,][]{1982MNRAS.198.1007,2003A&A...402..879O,
2006A&A...453..883V,2010NatPh...6..520P} incorporated some of these components. 
First simulations of ram pressure stripping including the dynamical effects of 
the magnetic fields were presented in \citet{2012arXiv1203.1343R} for 
late-type galaxies. 

We aim to systematically investigate how non-thermal components of the ISM and ICM affect ram pressure stripping in 
elliptical galaxies. This is our first paper in a series of papers on this subject, and it 
focuses on the effect of turbulent ISM on the ram pressure stripping rates, 
morphologies of the stripping tails, and mixing between the ISM and ICM. Although observational constraints on
the turbulence properties of the hot ISM in early-type galaxies are still uncertain, there is little doubt that the hot ISM 
is characterised by weak turbulence and randomly oriented weak magnetic fields 
\citep{2003ARA&A..41..191M,2004ARA&A..42..211E,2010MNRAS.402L..11S,2012arXiv1205.0256H}. 
Recent X-ray observations began to place meaningful constraints on the magnitude of turbulent motions in the hot gas of massive 
early-type galaxies, proving that the turbulence is subsonic 
\citep[e.g.,][]{2008MNRAS.388.1062C,2010MNRAS.404.1165C,
2010MNRAS.406..354O,2011MNRAS.410.1797S,2012A&A...539A..34D,2012ApJ...747...32B}. Stellar winds, supernovae, and active galactic 
nuclei are considered to be the main energy sources for these turbulent motions \citep{1996MNRAS.279..229M,2009ApJ...699..923B,2011ApJ...728..162D}.

We study the morphology of the ram pressure stripping tails. 
Sharp edges characteristic of ram pressure stripping have been detected in X-ray maps of a galaxy falling into the Fornax cluster
\citep{2005ApJ...621..663M}, and X-ray
tails are sometimes observed in ellipticals undergoing ram pressure stripping \citep[][]{2008ApJ...688..208R,
2008ApJ...688..931K, 2006ApJ...644..155M}. When the strength of stripping is not significant, 
or the duration of the process is short, galaxies are likely 
to show only somewhat elongated gas distributions instead of long tails \citep[e.g.,][]{2010MNRAS.405.1624M}.

In addition to the morphology of the gas distribution, ram pressure stripping also can be probed by tracking how well the 
ISM and ICM are mixed together for different ISM stripping rates \citep[e.g.,][]{2005A&A...435L..25S}. 
Most previous simulations focused on how material is 
expelled from galaxies into the ICM. Here we also examine how much mass can be mixed into galaxies from the ICM due to the 
random motions of the ISM in ram pressure stripping. 
A direct consequence of this mixing can be a significant change of the metallicity in the galactic gas.

The organisation of the paper is as follows. We describe the simulation setup in Section 2. In Section 3, we present 
results of our simulations, emphasising the differences in the impact of various strengths of turbulent motions on the
ram pressure stripping rates and tail morphologies. Finally, we present conclusions and discussion in Section 4.

\section{SIMULATIONS}

\subsection{Initial conditions}

Our galaxy model consists of dark matter halo and stellar mass distributions as well as 
the hot ISM. We assume that the gravitational field is dominated by the static distributions of the stellar and dark matter masses. 
Since galaxy cluster environments have strong tidal fields, we assume that the galactic gravitational field is truncated at the 
truncation radius $R_{t}=100$ kpc. Therefore, the gas experiences no 
gravitation acceleration once it escapes beyond $R_{t}$. 
The stellar mass distribution is described by a spherical Jaffe model, 
while the total mass distribution follows a $r^{-2}$ law \citep[e.g.,][]{2009MNRAS.393..491C,2009ApJ...699...89C,2010ApJ...711..268S}. 
Inside the effective radius of the stellar mass distribution, the mass of the dark matter halo 
is equal to the stellar mass. This setup has the effective radius of 3 kpc and the total stellar mass of ${\sim \rm 10^{11} M_{\odot}}$.

We setup the initial distribution of the ISM in hydrostatic equilibrium for the gravitational potential described 
above. However, this distribution is weakly perturbed by a stirring process which is explained in Section 2.3. 
Therefore, the precision of implementing the hydrostatic equilibrium condition is not critical in our initialisation. 
We note that satisfying the exact hydrostatic equilibrium condition is intrinsically difficult with finite 
volume methods with explicit time-stepping \citep{2002ApJS..143..539Z} because gravitational acceleration 
acts as a sink term in the momentum equation and can cause the growth of an instability \citep{Lian20101909}.

The initial ISM temperature profile has the following form 
\begin{equation}
T(r) = \left\{ \begin{array}{rl}
T_{i}                             &  \mbox{ if $r < r_{i}$} \\
2T_{0}/(1 + (r / r_{0})^{\beta})   &  \mbox{ otherwise}, 
\end{array} \right.
\end{equation}
where $T_{i}=8 \times 10^{6}$ K, $r_{i}=50.9$ kpc, $T_{0}=1.3\times 10^{7}$ K, $\beta=-3$, and $r_{0}=66.6$ kpc. 
The total mass of the ISM is ${\sim 4.4 \times 10^{10} M_{\odot}}$ inside $R_{t}$. 
The ICM is initialised with a constant density and temperature beyond $R_{t}$. 
The ICM temperature and density are equal to their corresponding ISM values at $R_{t}$.

We assume that ICM mean molecular weight and equation of state are same as those in the ISM. This assumption 
simplifies simulations by allowing us to use a single phase fluid. The temperature and density of the ICM are $2 \times 10^{7}$ K 
and $3 \times 10^{-28}$ ${\rm g ~ cm^{-3}}$, respectively.

The size of the simulation box is about 1150 kpc along $x$-axis which is the direction of the inflowing ICM. The length of 
the box is about 500 kpc in both $y$- and $z$-directions.  We make all zones cube-shaped 
by including more cells along the $x$-axis.

\subsection{Numerical methods}

We use the {\tt FLASH3} adaptive mesh refinement code with the most recent patches 
to solve Euler equations with gravity \citep{2000ApJS..131..273F,2009JCoPh.228..952L}. 
We employ a directionally unsplit staggered mesh hydro solver, Roe's approximate Riemann solver with van Leer flux limiter, 
the ideal gas equation of state with solar metallicity, and assume the gas to be fully ionised. 

We use two passive tracers -- a passive fluid and passive particles. 
The passive advection fluid (hereafter called ``colour'') is used to track the ISM mass fraction in cells. 
If a cell includes only the ISM, the value of this advection quantity is 1. 
Using this colour quantity we can measure mixing between two different materials 
\citep[e.g.,][]{2008ApJ...680..336S}. We can also determine where the mixed gas originated from by assigning different tag numbers to 
passive ISM and ICM particles that follow the fluid \citep{2009AnRFM..41..375T, 2007MNRAS.374..787H}. 
We distribute 8168 particles, which correspond to the ISM, uniformly within $R_{t}$ at the initial time.

We use the colour quantity as a refinement variable because we focus on resolving the mixing patterns 
of the ISM after it is stripped from the galaxy. 
Since there is no well-defined rule to choose specific refinement conditions or refinement variables 
\citep[][for discussion]{1989JCoPh..82...64B, 2010JCoAM.233.3139L}, 
we adopt the standard refinement method in the {\tt FLASH} 
code\footnote{\url{http://www.asci.uchicago.edu/site/flashcode/user_support/flash3_ug_3p3/node14.html#SECTION05163000000000000000}}. 
Regions exhibiting stronger 
variations in the passive scalar magnitude are more finely resolved. Specifically, we use {\tt refine\_cutoff}=0.8, {\tt derefine\_cutoff}=0.2, and
{\tt refine\_filter}$=10^{-2}$. 
The refinement level outside the truncation radius is allowed to vary between 3 and 6, but it is fixed at 5 for smaller distances. 
The maximum spatial resolution is 1 kpc outside $R_{t}$, and 2 kpc inside this radius.

\subsection{Stirring and inflow}

\begin{table}
\caption{Simulation runs}
\label{tab:run}
\begin{tabular}{@{}lcc}
\hline
Name & ISM injection energy & ISM 1D RMS velocity\\
& (${\rm cm^{2} ~ s^{-3}}$) & (Mach number) \\
\hline
Run 0 & $2.5 \times 10^{-8}$ & 0.022\\
Run 1 & $2.5 \times 10^{-7}$ & 0.038\\
Run 2 & $5.0 \times 10^{-7}$ & 0.048\\
Run 3 & $1.0 \times 10^{-6}$ & 0.067\\
Run 4 & $2.0 \times 10^{-6}$ & 0.093\\
Run 5 & $4.0 \times 10^{-6}$ & 0.119\\
\hline
\end{tabular}
\end{table}

We use a stirring module in the {\tt FLASH3} code 
\citep{1988CF.....16..257E,DBLP:journals/ibmrd/FisherKLDPCCCFPAARGASRGN08,2010ApJ...713.1332R} 
and modify it to restrict kinetic energy injection to the region inside $R_{t}$. We investigate six different 
strengths of turbulent motions as shown in Table \ref{tab:run}. The injection energies quoted in Table \ref{tab:run}
correspond to the energy per unit mass per mode. We consider 152 driving modes, and inject energy on scales between 49 to 50 kpc (see Appendix for possible effects of the injection scales). 
Our random forcing scheme uses stochastic Ornstein-Uhlenbeck process and a correlation timescale of 0.01 Gyr. 
These parameters result in turbulent motions with mass-weighted root-mean-square 
1D Mach numbers of approximately 0.022, 0.038, 0.048, 0.067, 0.093, and 0.119 for Run 0 to 5, respectively, 
before the onset of the ICM inflow at 0.5 Gyr.

The inflow velocity of the ICM is maintained at $\sim 170 {\rm km / s}$, which corresponds to 
Mach 0.25 with respect to the ICM sound speed. Even though galaxies in groups and clusters can move faster than the speed of sound, 
we note that the average speed for the entire orbit can vary between subsonic and supersonic depending on properties of 
galaxies and clusters. Moreover, velocities of early-type galaxies are likely to be biased toward lower values than those of spiral galaxies 
\citep[e.g.,][]{1998A&A...331..439A,2008ApJ...676..218H}. 
Here, we focus on the subsonic case as a simple model. If there is no obstacle, it takes about 6.6 Gyr 
for this flow to cross the whole simulation domain along the $x$-axis. 
All boundaries, except for the low-x (inflow) boundary, are outflow boundaries. We also test 
nine times higher ram pressure in Run 0 and 5 by adopting the inflow velocity of Mach 0.75. These simulations 
(hereafter, Run 0h and 5h) allow us find out how strongly the effects of the turbulent ISM change depending 
on the strength of the ram pressure.

We consider two different stirring cases. In Case A, we continuously stir the ISM with the constant 
injection energy as explained above. This case corresponds to the situation when the turbulent energy sources,  
such as active galactic nuclei or supernova explosions, continue to operate inside the galaxies. In Case B, the turbulent 
energy injection is stopped after 0.5 Gyr, at which point the inflow of the ICM begins. All simulations run up to 6 Gyr.

\section{RESULTS}

We performed twelve simulations for six different strengths and 
two different durations of the turbulent energy injection, that is, Cases A and B. We also simulated four strong ram pressure stripping 
cases for two strengths of turbulence driving in Case A and B. We now present the results of these simulations, focusing on relative differences among the runs. 

\subsection{Overall evolution}

\begin{figure*}
\includegraphics[width=165mm]{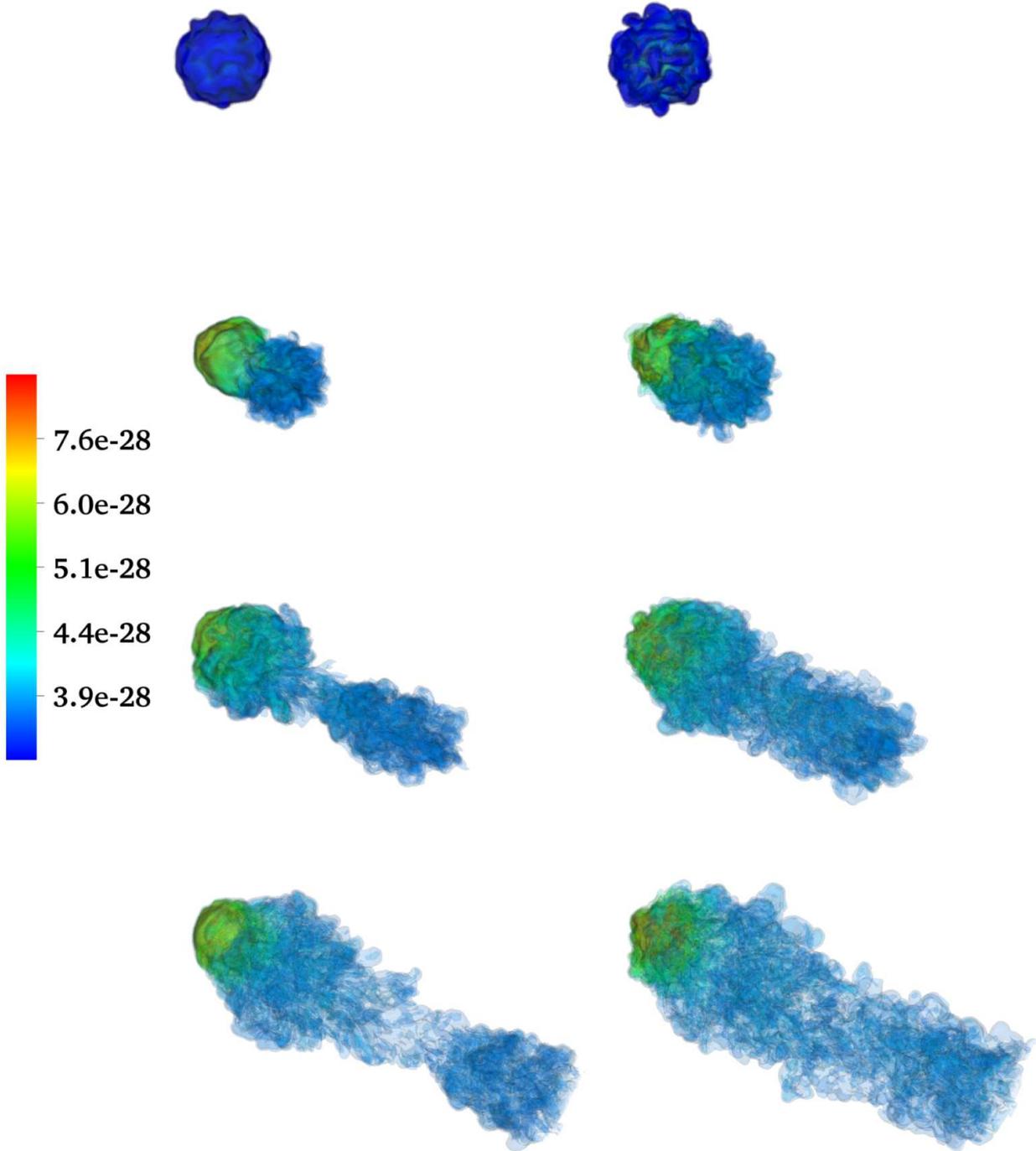}
\caption{ISM density distributions for ten isosurfaces of the ISM mass fractions ranging linearly from 0.01 to 1. 
The unit of density is ${\rm g\; cm^{-3}}$. Runs 1 and 4 for Case A are shown in the left and right columns, 
respectively. From top to bottom, each row corresponds to 0.75, 2, 4, and 6 Gyr. Because the top two panels show 
the early phase, the radius of the approximately spherical ISM distribution is close to $R_{t}=100$ kpc.}
\label{fig:3D}
\end{figure*}

Figure \ref{fig:3D} shows the time sequence of the evolution of the ISM in a galaxy that is subject to the ram pressure stripping.
Left column corresponds to Run 1 and right one to Run 4. Run 0 
corresponds to extremely weak stirring and can be thought of as a reference non-turbulent case. Both columns are for Case A, where the 
stirring energy is continuously supplied to the ISM. From top to bottom, each row corresponds to 0.75, 2, 4, and 6 Gyr.
This figure demonstrates that the turbulence is well developed before the ICM wind begins to interact with the galaxy.
The ram pressure produces a long turbulent tail, and the tail properties vary with the strength of the ISM turbulence.
In particular, Run 1 reveals a more discontinuous tail than Run 4. This difference becomes more evident 
at later times. Moreover, Run 4 corresponds to a longer and broader tail than Run 1.
This is due to the fact that the stripping from the outer ISM layers is enhanced in Run 4. We quantify these morphological 
differences in the following subsection.

\begin{figure*}
\includegraphics{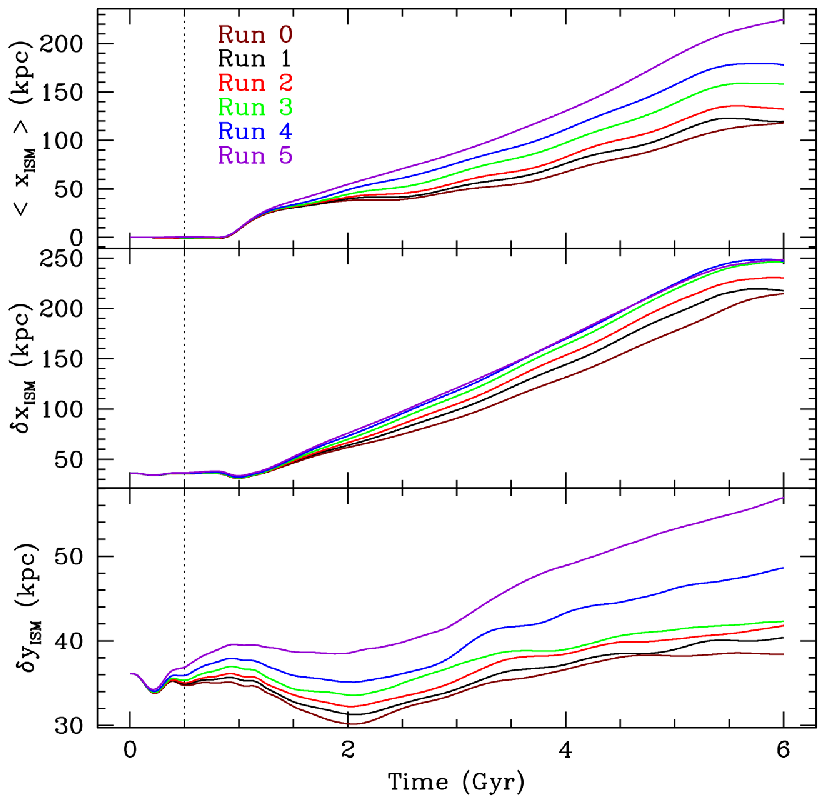}
\includegraphics{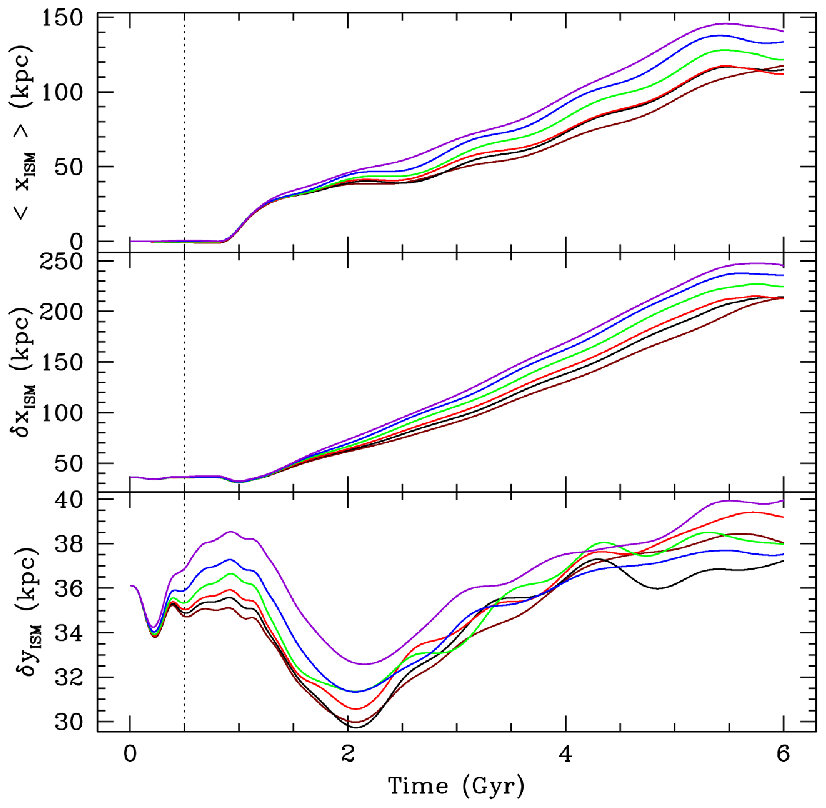}
\caption{Evolution of ${\rm\langle x_{ISM}\rangle}$,  ${\rm \delta x_{ISM}}$, and ${\rm \delta y_{ISM}}$. 
The simulations with the continuous supply of the turbulent energy  ({\it left}; Case A) lead to 
larger differences in ${\rm\langle x_{ISM}\rangle}$ among the runs than those without the continuous energy supply ({\it right}; Case B). 
The same tendency is found in the distributions of ${\rm \delta x_{ISM}}$ and ${\rm \delta y_{ISM}}$. }
\label{fig:color_global}
\end{figure*}

\subsection{Tail morphology}
In order to quantify the evolution of the spatial distribution of the ISM, we use colour weighting (see above).
The colour quantity $C$ represents the ISM fraction in cells, and we use it to obtain the mass of the ISM in each cell. 
We then employ the following equations to describe the evolution of the tail properties

\begin{equation}
\langle x_{\rm ISM}\rangle = \frac{\sum_{i} C_{i} \rho_{i} V_{i} x_{i}}{\sum_{i} C_{i} \rho_{i} V_{i}},
\label{eq:xavg}
\end{equation}
\begin{equation}
\delta x_{\rm ISM} = \sqrt{ \frac{\sum_{i} C_{i} \rho_{i} V_{i} ( x_{i} -\langle x_{\rm ISM}\rangle )^{2}}{\sum_{i} C_{i} \rho_{i} V_{i}} },
\label{eq:xstd}
\end{equation}
where the index $i$ represents cell number, and $\rho$, $V$, and $x$, correspond to density, cell volume, and cell $x-$coordinates, respectively. 
The quantity ${\rm\langle x_{\rm ISM}\rangle}$ traces the overall shift of the ISM mass after it is stripped from the galaxy. 
Finally, ${\rm \delta x_{\rm ISM}}$, and its $y-$ direction counterpart ${\rm \delta y_{\rm ISM}}$, quantify the widths the distributions of the ISM along the $x$ and $y$ axes.

\begin{figure*}
\includegraphics[scale=0.43]{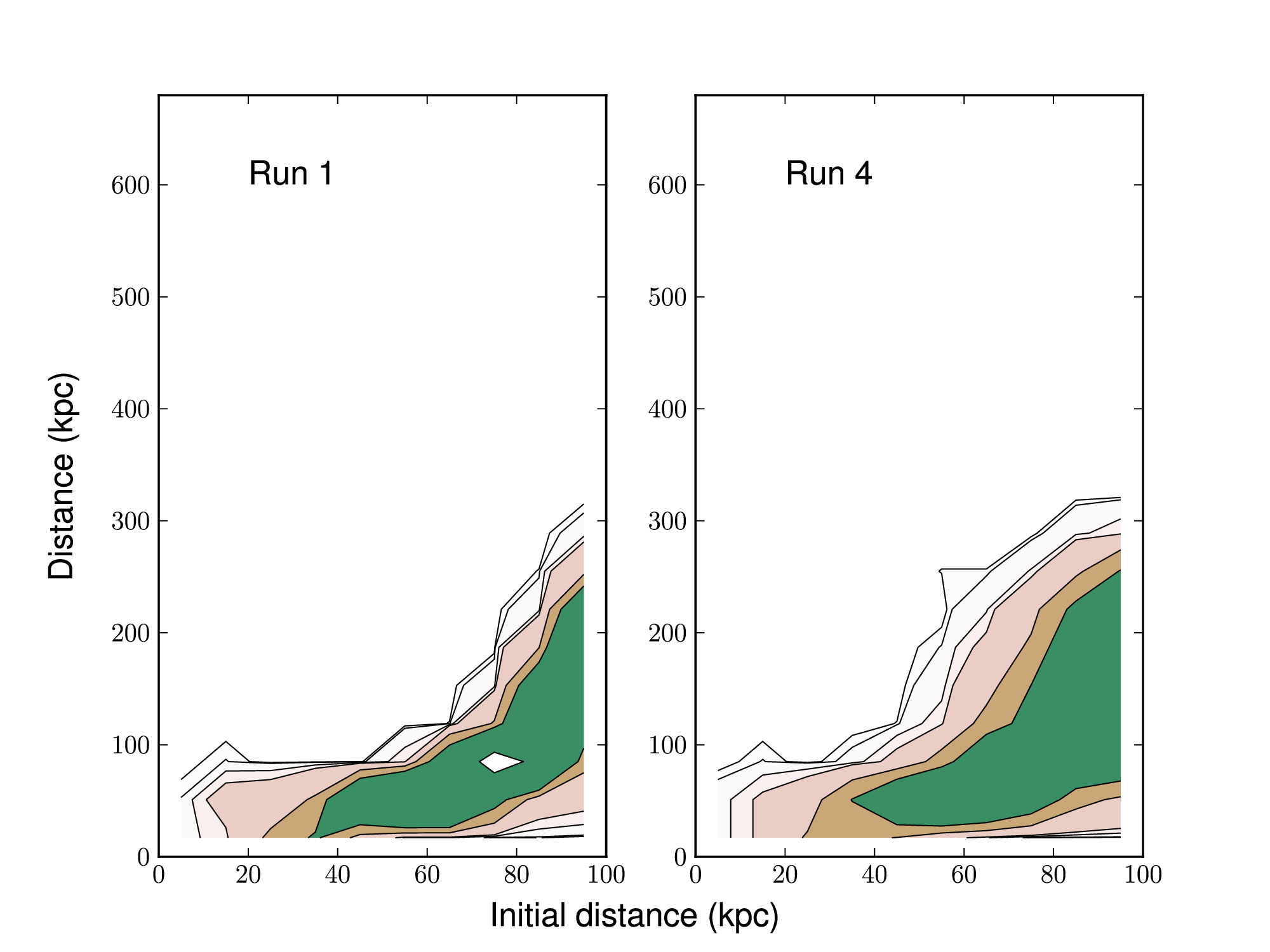}
\includegraphics[scale=0.43]{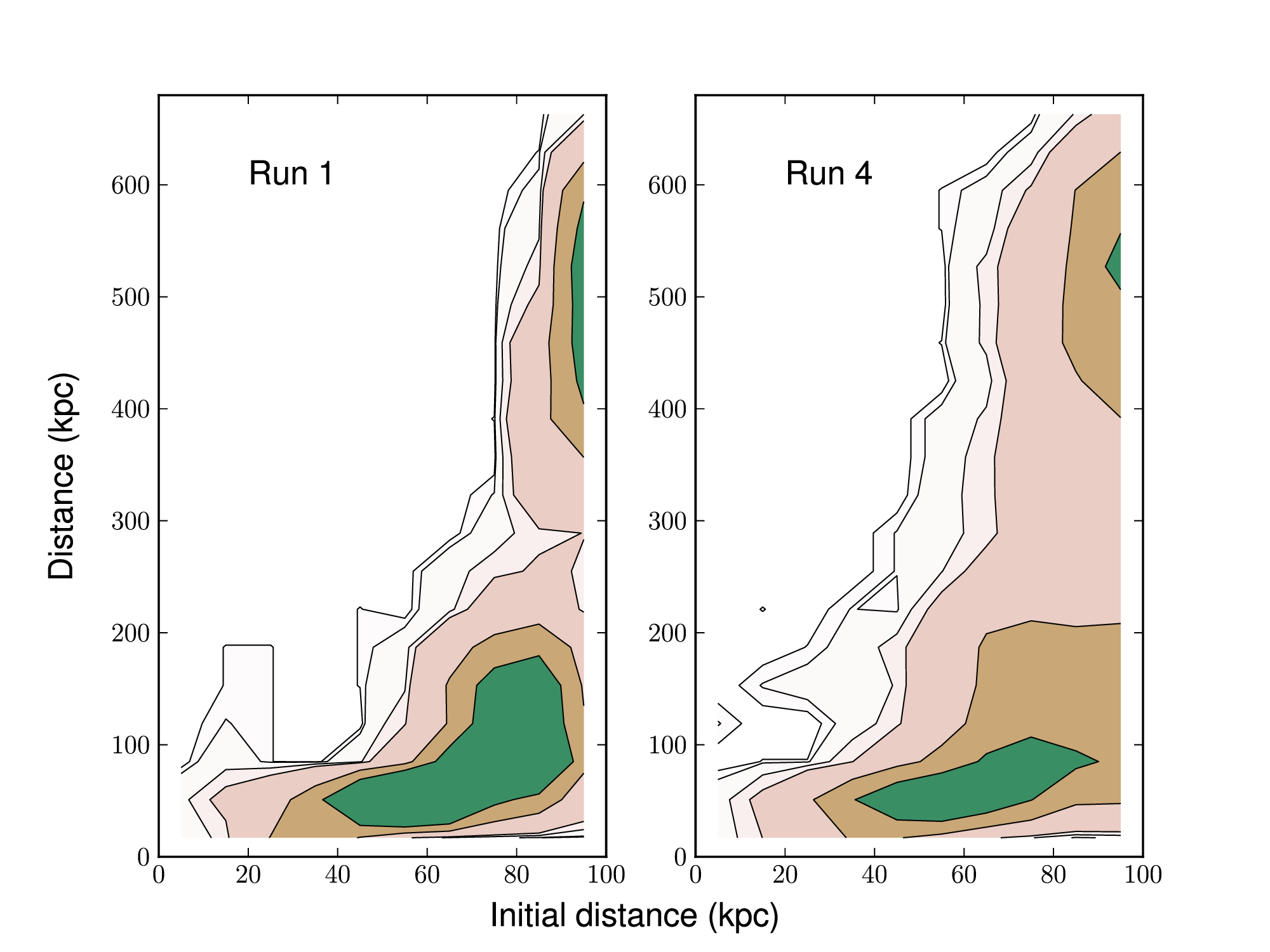}
\caption{Density distributions of the passively moving particles in Runs 1 and 4 at 2 {\it (left)} and 4.5 {\it (right)} Gyr. 
The number density of the passive particles is measured on uniform grids with bin sizes of 10 and 34 kpc for horizontal and vertical axes, respectively. 
The contour lines correspond to the number densities $5 \times 10^{-5}$, $1 \times 10^{-4}$, $5 \times 10^{-4}$, $1 \times 10^{-3}$, 
$5 \times 10^{-3}$, $1 \times 10^{-2}$, and $4 \times 10^{-2} ~ {\rm kpc^{-2}}$. All plots show the distributions corresponding to Case A.}
\label{fig:particle_global}
\end{figure*}

Figure \ref{fig:color_global} shows the evolution of ${\rm\langle x_{\rm ISM}\rangle}$, ${\rm \delta x_{\rm ISM}}$, and 
${\rm \delta y_{ISM}}$. These measurements quantify what is shown in Figure \ref{fig:3D}. 
When the stirring process is continuous (Case A; left panel), 
Run 4 creates a longer tail than Run 1, and this difference increases with time up to 
60 kpc at 6 Gyr. However, the difference is much smaller in  Case B runs (right panel). 
This comparison also confirms that the initial expansion of the ISM caused by the injected energy from the 
stirring process is not the cause for the difference between Runs 1 and 4 in Case A. If the initial expansion 
was the main reason for the differences among different runs in Case A, the same effect should be 
found in Case B too. Yet, in the absence of the continuous stirring (Case A), no significant differences 
are seen among different runs. Therefore, the continuous supply of the turbulent energy causes differences found 
between Cases A and B.

\begin{figure*}
\includegraphics[scale=1.0]{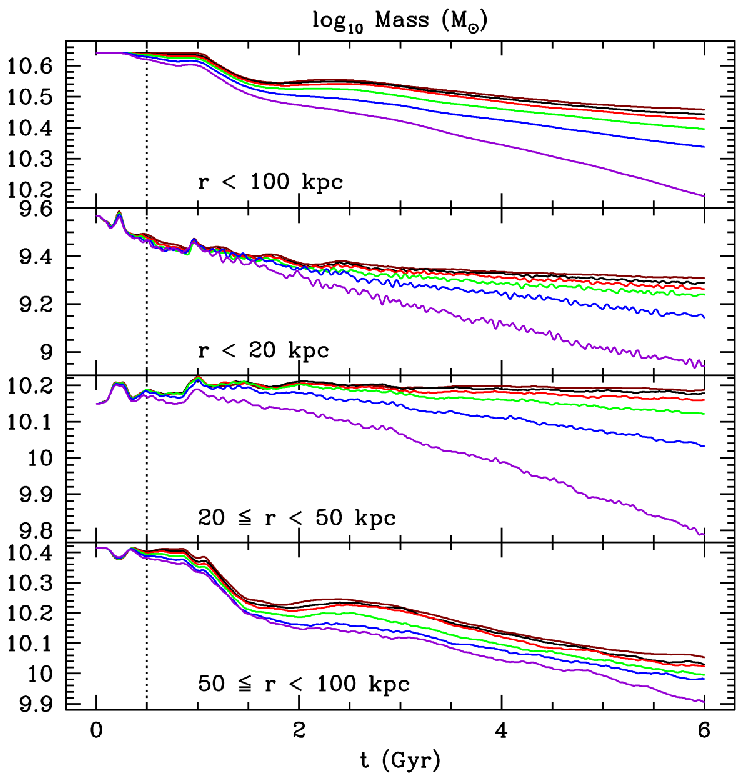}
\includegraphics[scale=1.0]{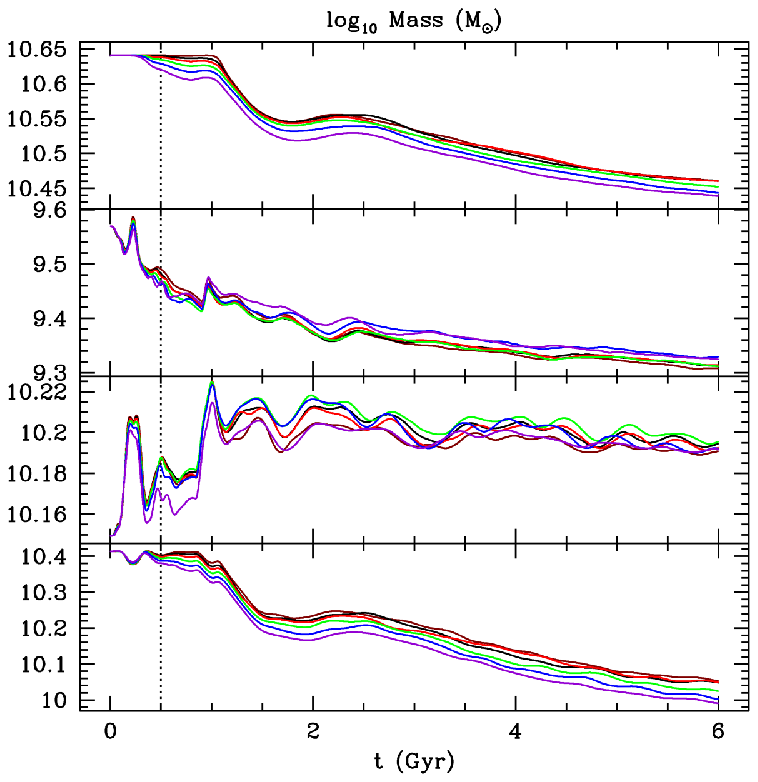}
\caption{Mass evolution of the intrinsic ISM for Case A ({\it left}) and Case B ({\it right}). 
From top to bottom, each panel shows the mass in four different radial zones:  $r < 100$ kpc, 
$r < 20$ kpc, 20 kpc $\leq r <$ 50 kpc, and 50 kpc $\leq r <$ 100 kpc 
($R_{t}=100$ kpc). The dotted line corresponds to 0.5 Gyr when the inflow of the ICM starts to enter the simulation box. 
The colour coding of the different lines is the same as in Figure \ref{fig:color_global}. 
We note that the ranges of the vertical axes are not same in the left and right columns.}
\label{fig:mass_color_ISM}
\end{figure*}

\begin{table}
\caption{Times corresponding to the removal of 
10\%, 20\%, and 30\% of the ISM in Case A (measured from the onset 
of the ICM inflow at 0.5 Gyr.}
\label{tab:time_scale}
\begin{tabular}{@{}lccc}
\hline
Name & ${\rm \Delta t_{10}}$ (Gyr) & ${\rm \Delta t_{20}}$ (Gyr) & ${\rm \Delta t_{30}}$ (Gyr) \\
\hline
Run 0 & 0.746 & 2.314 & 4.010 \\
Run 1 & 0.729 & 1.192 & 3.721 \\
Run 2 & 0.709 & 1.079 & 3.433 \\
Run 3 & 0.672 & 0.977 & 2.815 \\
Run 4 & 0.619 & 0.896 & 2.132 \\
Run 5 & 0.546 & 0.789 & 1.242 \\
\hline
\end{tabular}
\end{table}

Figure \ref{fig:color_global} shows that stronger turbulence in Case A results in wider dispersions of the ISM  
parallel and perpendicular to the direction of the ram pressure. Even though in Case B the energy 
is not supplied continuously to the ISM, the evolution of ${\rm \delta x_{ISM}}$ is similar to what we find in Case A.
We do not find a significant effect of turbulence on ${\rm \delta y_{ISM}}$ in Case B. 
Although Case B does not exhibit significant deviations among different runs, general trends in the tail evolution 
Case B is the same as in Case A. 
In particular, at around 2 Gyr, the initial stripping generates short and narrow tail immediately behind the galaxy 
(see Figure \ref{fig:3D}). This is due to a converging ICM flow behind the galaxy. 
This narrowing is a transient feature and the tail widens after 2 Gyr.

\subsection{Origin of the ISM in the tails}

The origin of the ISM stripped away from the galaxy can be used to understand what kind of materials are transported to the ICM 
by the ram pressure stripping. 
As explained in the previous section, we map the
initial positions of the particles to their temporal positions. This allows us to check where exactly the stripped material in the tail came from.  

Figure \ref{fig:particle_global} shows the distributions of particle surface densities in the two-dimensional space defined by 
the initial and current particle positions measured with respect to the galactic centre. 
Left panel corresponds to 2 Gyr and the right one to 4.5 Gyr. 
Initially, all particles were distributed uniformly inside the truncation radius.

The results presented in Figure \ref{fig:particle_global} allow us to make to points. 
First, most of the stripped gas originates from the outer parts of the galaxy
near the truncation radius, and, second, higher turbulence levels enable gas removal from deeper layers of the galactic atmosphere.
Regarding the first point, the gas originally located near the truncation radius is always transported to largest radii, 
independently of the turbulence level.
Regarding the second point, a significant amount of the ISM initially residing between 60 and 80 kpc is stripped beyond 200 kpc at 2 Gyr in Run 4. 
This makes the 
distribution of the particle density in Run 4 appear wider at a given distance beyond 200 kpc than in Run 1.
This is again because more vigorous ISM turbulence in Run 4 more efficiently transports the ISM from small radii to the ISM-ICM
interface from which the gas is permanently removed.

Figure \ref{fig:particle_global} also shows that the narrow part of the tail found in Run 1 (see Figure \ref{fig:3D}) is caused by
inefficient stripping of the ISM. At 4.5 Gyr, Run 1 has fewer particles at 
$\sim$300 kpc than Run 4, which leads to narrower tail. This low-density structure of 
the tail is caused by the inefficient stripping of the ISM that initially resided at radii larger than 60 kpc.

\subsection{Evolution of the ISM mass retained in the galaxy}

Since we use the passive scalar quantity to identify ISM and ICM separately, 
we can follow the evolution of the gas that originally belonged to the ISM. 
Figure \ref{fig:mass_color_ISM} shows the evolution of the ISM mass in four different radial bins inside ${R_{t}}$.
We find significant differences between Case A and B. A continuous supply of the turbulent 
energy enhances internal mixing of the ISM, resulting in the increase in the net mass loss rate of the ISM. 
This effect is not seen in Case B.

We find that the strength of turbulence has a noticeable effect on the distribution of the intrinsic ISM inside ${\rm R_{t}}$ in Case A.
As shown in Figure \ref{fig:mass_color_ISM}, in Case A 
at 6 Gyr, Run 0 retains about ${\rm 2.87 \times 10^{10} ~ M_{\odot}}$ of the intrinsic ISM, 
while Run 5 has about ${\rm 1.51 \times 10^{10} ~ M_{\odot}}$, i.e. about 2 times less. 
\ref{tab:time_scale} summarises time scales of 10, 20, and 30\% ISM mass loss after the inflow of ICM produces ram pressure 
in Case A. 
In Case B, although we find the same general trend as Case A, the difference is smaller.

\begin{figure}
\includegraphics[scale=0.43]{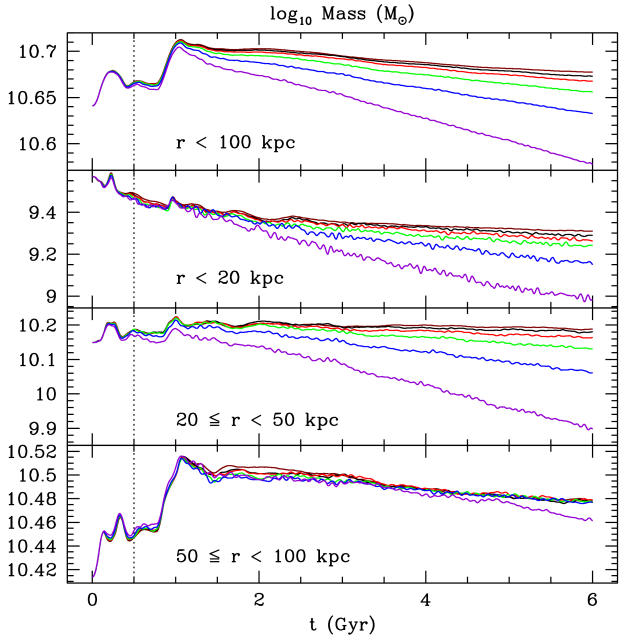}
\caption{Evolution of the total gas mass inside $R_{t}=100$ kpc for Case A. 
From top to bottom, each panel shows the mass in four different radial zones:  $r <$ 100 kpc, $r <$ 20 kpc, 20 kpc $\leq r <$ 50 kpc, and 
50 kpc $\leq r <$ 100 kpc. The dotted line corresponds to 0.5 Gyr when the inflow of the ICM starts to enter the simulation box. 
The colour coding of the different lines is the same as in Figure \ref{fig:color_global}.}
\label{fig:mass_ISM}
\end{figure}

\begin{figure}
\includegraphics[scale=0.43]{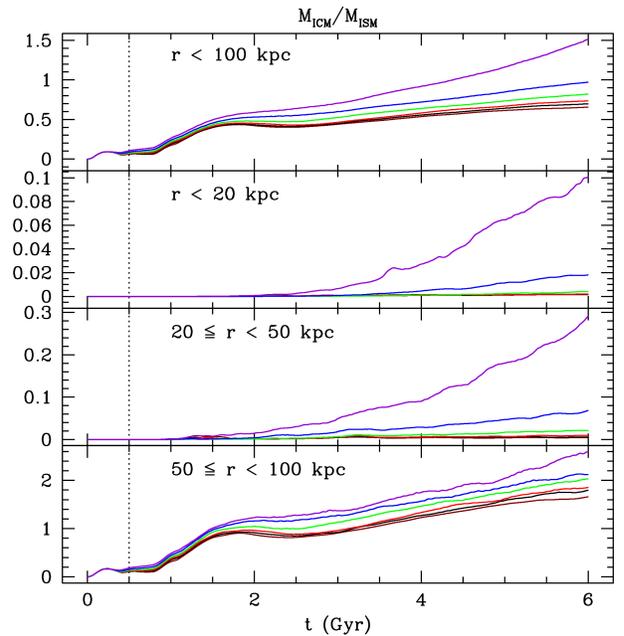}
\caption{Evolution of the ratio of the intrinsic ICM mass over the intrinsic ISM mass inside ${R_{t}}$ for Case A. 
The colour scheme is the same as in Figure \ref{fig:color_global}.}
\label{fig:ICM_over_ISM}
\end{figure}

\begin{figure}
\includegraphics[scale=0.43]{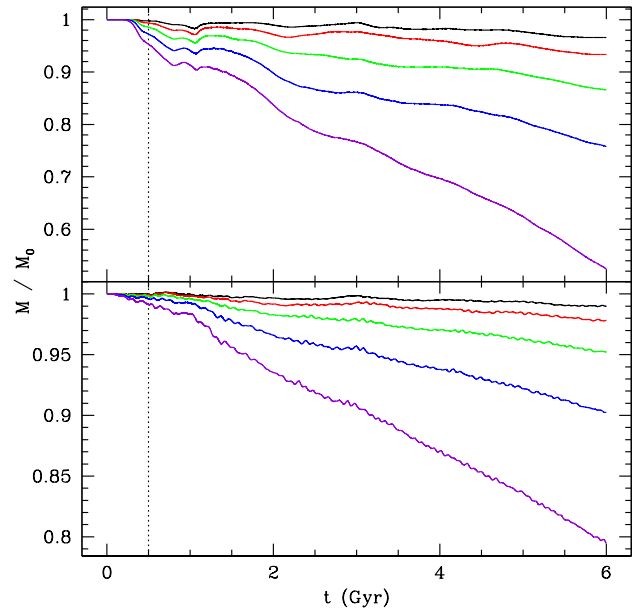}
\caption{Evolution of the intrinsic ISM mass ({\it top}) 
and the total gas mass ({\it bottom}) inside ${R_{t}}$ with respect to masses for Run 0 in Case A. 
The colour scheme is the same as in Figure \ref{fig:color_global}.}
\label{fig:mass_ISM_ICM}
\end{figure}

This evolution of the intrinsic ISM mass hints at the possibility that, in Case A, a significant amount of 
the inflowing ICM penetrates the galaxy and mixes with the ISM. 
The volume initially occupied by the intrinsic ISM can be partially refilled by the inflowing ICM 
if the ISM-ICM mixing is efficient within ${R_{t}}$.

\subsection{Evolution of the total gas mass inside the galaxy}

In Figure \ref{fig:mass_ISM}, we show the evolution of the total gas mass including both the ISM and ICM inside $R_{t}$. 
We find that the ICM temporarily accumulates mainly over $50 \leq r < 100$ kpc in Case A. The total mass of the gas increases up to ${\rm \sim 
3.3 \times 10^{10} ~ M_{\odot}}$ over $50 \leq r < 100$ kpc as the ICM compresses the ISM and then blends with the ISM. However, the ICM caught in 
the galaxy is finally expelled after $\sim$1 Gyr by the combined action of the ram pressure stripping and continuous supply of turbulent energy.

As shown in Figure \ref{fig:mass_color_ISM}, the mass of the intrinsic ISM does not depend sensitively on the level of turbulence in Case B.
Similarly, for the total gas mass within $R_{t}$, we do not observe strong trends with the turbulence strength in Case B, and therefore we do not
show Case B in  Figure \ref{fig:mass_ISM}. However, overall evolution of the total mass as a function of radius in Case B is similar to that in Case A.

\begin{figure*}
\includegraphics{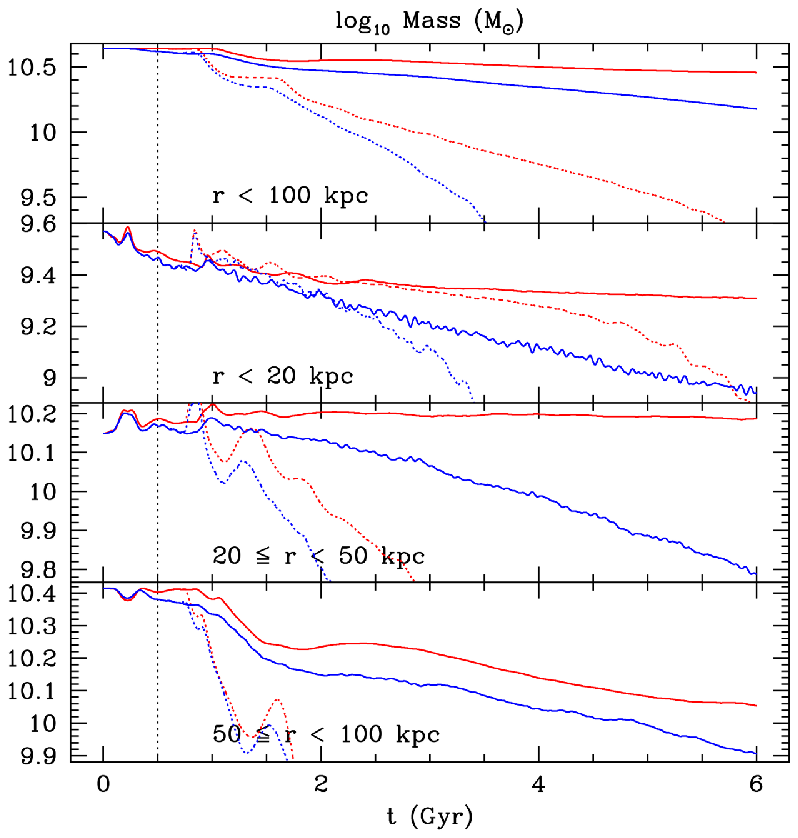}
\includegraphics{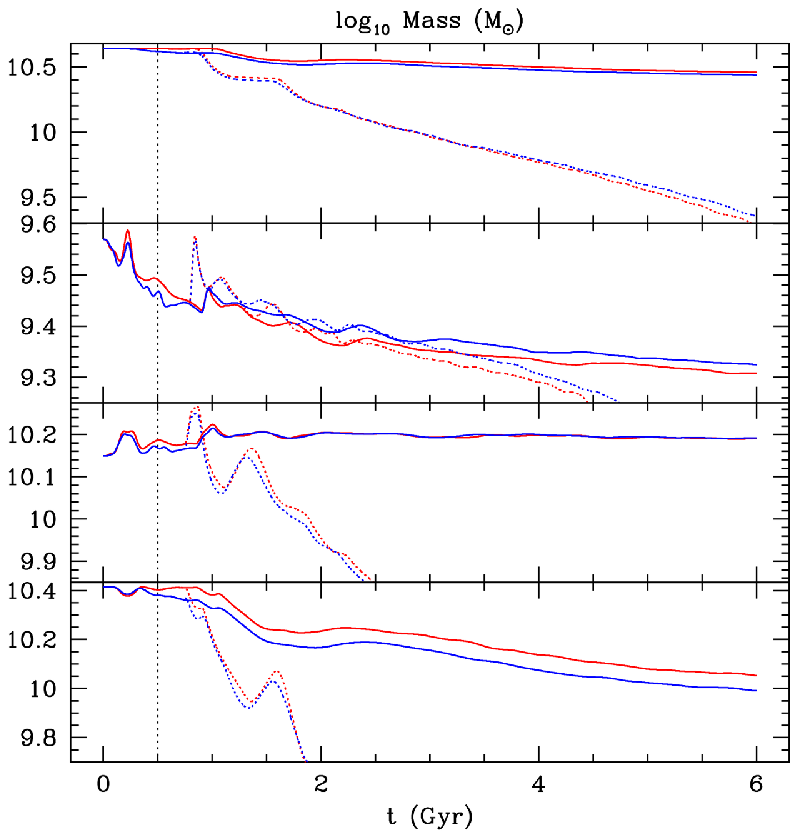}
\caption{Mass evolution of the intrinsic ISM for Case A ({\it left}) and Case B ({\it right}) in Run 0 ({\it red}) and 
5 ({\it blue}) for the low ({\it solid line}) and high ({\it dotted line}) ram pressure stripping. 
From top to bottom, each panel shows the mass in four different radial zones:  $r < 100$ kpc, 
$r < 20$ kpc, 20 kpc $\leq r <$ 50 kpc, and 50 kpc $\leq r <$ 100 kpc 
($R_{t}=100$ kpc). The dotted vertical line corresponds to 0.5 Gyr when the inflow of the ICM starts to enter the simulation box. 
We note that the range of the vertical axe are not same in the left and right columns for the panel of 20 kpc $\leq r <$ 50 kpc.}
\label{fig:mass_color_ISM_high_speed}
\end{figure*}

The fractional change in the total gas mass within $R_{t}$ is smaller than the fractional change in the intrinsic ISM retained within $R_{t}$ 
(c.f. top panel in Figure \ref{fig:mass_ISM} and top left panel in Figure \ref{fig:mass_color_ISM}, respectively). This suggests that the ICM 
mixes with the intrinsic ISM.

Figure \ref{fig:ICM_over_ISM} summarises how much mass is contributed by the ICM and ISM inside $R_{t}$ as a function of time for Case A. 
As the strength of the turbulent motions in the ISM increases, the fraction of the ICM penetrating into the galaxy increases, 
resulting in about 1.5 times more ICM mass than the ISM mass inside  $R_{t}$ in Run 5 at 6 Gyr. Considering only the outer region of the galaxy over 
$50 \leq r < 100$ kpc, this fraction is about 2.6 in Run 5 while it becomes about 1.6 in Run 0.

Figure \ref{fig:mass_ISM_ICM} presents the relative differences in the mass loss caused by different strengths of the ISM turbulence.
At 6 Gyr, Run 5 retains almost twice less ISM than Run 0, which corresponds to weak ISM 
turbulence. Yet, considering the total mass including ISM and ICM inside $R_{t}$, Run 5 has only 
20\% less mass than Run 0, because more ICM is blended with the ISM in Run 5  than in Run 0. 
Importantly, the differences among different runs show strongly non-linear dependence on the ISM velocity dispersion. 
In Run 4, the intrinsic ISM mass is about 25\% less than in Run 0 at 6 Gyr, 
while twice less intrinsic ISM is retained in Run 5 than in Run 0.

\subsection{Ram pressure stripping for higher ICM inflow velocity}

By comparing results from Run 0 and 5 to those from Run 0h and 5h, we investigate how the increasing strength of 
the ram pressure alters the effects of the turbulent ISM on the stripping efficiency. Figure \ref{fig:mass_color_ISM_high_speed} 
shows that the increased ram pressure of the ICM enhances stripping in both Runs 0 and 5 
Because the ram pressure is nine times stronger in 
Run 0h than in Run 0, the mass of the ISM left inside the galaxy is much lower in Run 0h than in Run 0. 
Similarly, Run 5h retains $\sim$13 times less gas than Run 5 at $\sim$3.5 Gyr.
The increased inflow speed significantly increases the efficiency of the initial stripping in the outer 
50 kpc $\leq r <$ 100 kpc region. As the stripping process continues, some amount of the ISM 
originally located at the centre of the galaxy (i.e.,  $r < 20$ kpc) is moved to the outer regions 
gradually by the turbulent ISM, and then is stripped from the galaxy. 
Therefore, the increase in the stripping efficiency is larger in Run 5h than 0h.
For example, top left panel in Figure \ref{fig:mass_color_ISM_high_speed} shows 
that Run 0h retains $\sim$3 times more intrinsic ISM at $\sim$3 Gyr since the onset of 
stripping than Run 5h.

Comparison of Cases A and B in Figure \ref{fig:mass_color_ISM_high_speed}, shows that 
continuous supply of the turbulence energy amplifies the efficiency of ram pressure stripping.
In Run 5h, Case A keeps about 4 times less ISM than Case B at 
around 3.5 Gyr. The difference between Case A and B is particularly striking in the central regions ($r < 20$ kpc).
We note that we continuously supply the turbulence energy in Case A even after the substantial 
amount of the ISM is stripped at around 2 Gyr. This might not be a realistic assumption for turbulence driven by 
stellar and/or AGN processes.
After the large amount of the cold ISM has been removed from a galaxy, the galaxy may not be able to generate strong turbulence due to 
processes such star formation, supernova explosions, and AGN feedback. 
Consequently, the strong ram pressure stripping results beyond $\sim$3 Gyr may not be reliable.
However, we argue that these cases serve to bracket the range of possible solutions.

\section{Discussion and Conclusions}

We show that the continuous supply of small to moderate amount of turbulent kinetic energy to the ISM enhances 
the ISM mass loss rate in elliptical galaxies experiencing ram pressure stripping, and 
increases the penetration of the ICM into the galaxies (see Figure \ref{fig:mass_ISM_ICM}). 
The spatial distribution of the stripped ISM can be wider and more extended along the 
direction of galaxy motion (see Figures \ref{fig:3D}, \ref{fig:color_global}, and 
\ref{fig:particle_global}), when AGN feedback and/or stellar processes such as star formation are present. 
Our results imply that early-type galaxies 
characterised by the turbulent ISM should efficiently disperse their ISM throughout galaxy clusters.

The origin of the stripped ISM in the tails shows that the ram pressure stripping with the turbulent motions in the ISM boosts the mixing between 
the central region and the outer region of the galaxy. This implies that the distributions of gas properties in the tails can be 
used to infer the distribution of the intrinsic ISM properties such as gas metallicity inside galaxies. 
Since the distant part of the tail in Run 5 is more mixed with the central gas inside the galaxy than 
in Run 0, we expect that the properties of the stripped ISM should show weaker gradients of the gas properties 
along the tail in Run 5 than in Run 0.

For example, there might be a gradient in metallicity distribution along the ram pressure stripping tail. 
In general, the ISM in the central regions of early-type galaxies is more metal-rich than in the outer regions \citep[e.g.,][]{2011MNRAS.418.2744M,
2011ApJ...729...53H}. 
Therefore, very low levels of the ISM turbulence in early-type galaxies make the stripping tail have only low metallicity ISM 
along the tail, contributing negligibly to the ICM 
metal enrichment \citep[see,][for discussion of the ICM enrichment efficiency]{2008SSRv..134..363S,2008ApJ...688..931K}. However, we 
note that this depends on the initial metallicity distribution in the galaxy. If 
a galaxy has a shallow metallicity gradient before experiencing ram pressure stripping, it can flatten the 
metallicity distribution along the tail and lead to more significant ICM enrichment even when the strength of the ISM turbulence is weaker. 

The evolution of the mass inside the galaxy implies 
that a significant fraction of the gas mass measured in observations can be explained by the ICM gas 
that got temporarily incorporated into the ISM. 
As Figure \ref{fig:ICM_over_ISM} shows, galaxies with the strong 
turbulent motions in the ISM easily blend the inflowing ICM with the ISM. 
Therefore, the properties of the hot X-ray emitting ISM in the galaxy 
experiencing ram pressure stripping might have been altered by the inflowing ICM, in particular, in the 
outer regions.

For example, the metallicity of the ICM in low-redshift galaxy clusters is about 0.5 $Z_{\odot}$ \citep{2009ApJ...698..317A} 
and can be much lower than that of the ISM \citep{2006ApJ...639..136H,2009ApJ...696.2252J}, 
Thus, mixing of the ICM with galactic gas can alter the metallicity of the ISM and the metallicity 
gradient inside cluster galaxies. This contamination can be particularly significant in the outer regions of galaxies when the ISM is turbulent. 
Interestingly, if the galaxy has an initially flat metallicity profile at the level of 2 $Z_{\odot}$, 
and if the metallicity of the ICM is about 0.5$Z_{\odot}$, then the
ram pressure stripping will lower the mass-weighted metallicity to around 0.9 and 1.9 $Z_{\odot}$ in 
50 $\leq r <$ 100 kpc and $r <$ 20 kpc, respectively (with the mass ratios shown in Figure \ref{fig:ICM_over_ISM} for Run 5 at 6 Gyr) 
This specific case illustrates how ram 
pressure stripping in the presence of turbulent ISM can steepen ISM metallicity profiles. This steepening effect 
is expected to be more pronounced as the strength of ISM turbulence increases.

The models presented here allow one to study the effects of turbulence on the ram pressure stripping process
via a conceptually simple approach. The advantage of this approach lies in providing a clear intuitive picture of 
how the turbulent ISM affects the gas stripping. 
These models form a framework for future studies that will relax some of the assumptions made in the present work.

Our current simulations do not include a few important physical processes that are required to make detailed observational predictions 
for the ram pressure stripping process. 
First, we do not include radiative cooling processes \citep[see,][for a review]{2008SSRv..134..155K}, 
which will lead to the formation of dense cold gas clouds \citep[e.g.,][]{2007ApJ...671..190S,2010ApJ...717..147S,2010ApJ...722..412Y}. 
Second, self-gravity of the gas is not included. Self-gravity can alter the evolution of 
the stripped ISM by accelerating the collapse of these dense cold gas clouds. 
Third, the spatial resolution of our simulations is not high enough to fully cover an extremely broad inertial range of the turbulent ISM 
\citep[see][for a review]{1998AnRFM..30..539M,2011RPPh...74d6901B}. 
Fourth, we have neglected magnetic fields, which may affect the efficiency of mixing of the ISM and ICM, suppress viscosity and thermal conduction 
between the stripping tail of the cold gas and the hot ICM, and introduce non-trivial dynamical effects. 
Finally, the energy sources of the turbulence in our simulations are not directly controlled by the relevant 
astrophysical processes such as star formation and AGN. 
Continuous mass loss by ram pressure stripping can affect star formation and AGN feedback 
\citep{1992MNRAS.255..346B,1999MNRAS.309..161M,2008A&A...481..337K,2012ApJ...745...13S}, increasing or decreasing energy injected 
to the turbulent ISM.

Our main concern is the fact that our model currently does not take into account a possible coupling 
between the efficiency of stirring of the gas by star formation and AGN and the efficiency of stripping. For example, it is conceivable that 
enhanced stellar or AGN feedback could increase the level of turbulence, accelerate the mass removal from the galaxy, and thus  
reduce the fuel supply for these feedback processes and the efficiency of the ram pressure stripping process. 
Consequently, less gas would be available to fuel AGN and star formation, and the stirring efficiency would slow down. 
Our model currently does not incorporate such mechanism. However, we show that the efficiency of ram pressure stripping depends 
sensitively on the duration of stirring, and our models for the continuous (Case A) and 
initial (Case B) stirring likely bracket the range of possibilities.

In future work, we will relax some of the assumptions and simplifications made here. The second paper 
in this series will investigate the effect of weakly magnetised turbulent ISM in elliptical galaxies 
on the ram pressure stripping process.

\section*{Acknowledgements}

We are grateful to Karen Yang and Dongwook Lee for useful discussions. 
We thank the referee (Eugene Churazov) for his valuable comments that improved this manuscript. 
MR acknowledges NSF grant 1008454. 
This work used the Extreme Science and Engineering Discovery Environment (XSEDE), 
which is supported by National Science Foundation 
grant number OCI-1053575. The software used in this work was in part developed by 
the DOE NNSA-ASC OASCR Flash Center at the University of Chicago.

\appendix

\section{Effect of turbulence driving scales}

The properties of the ISM turbulence can affect the stripping efficiency. In particular, 
the stripping efficiency can depend on the outer turbulence driving scale $l_{turb}$. 
This is expected because the effective diffusion coefficient $\sim ~ l_{turb} \, v_{turb}$, 
where $v_{turb}$ is a characteristic turbulence velocity. In order to investigate 
this dependence, we perform an additional simulation which corresponds to Run 5 for Case A, 
but where the outer turbulence scale is reduced from $\sim$ 50 kpc to $\sim$ 25 kpc, and 
where the energy injection rate per mode is adjusted such that the total amount of the 
turbulent energy injection within $R_{t}$ is the same as in the original Run 5 for Case A.

\begin{figure}
\includegraphics[scale=0.43]{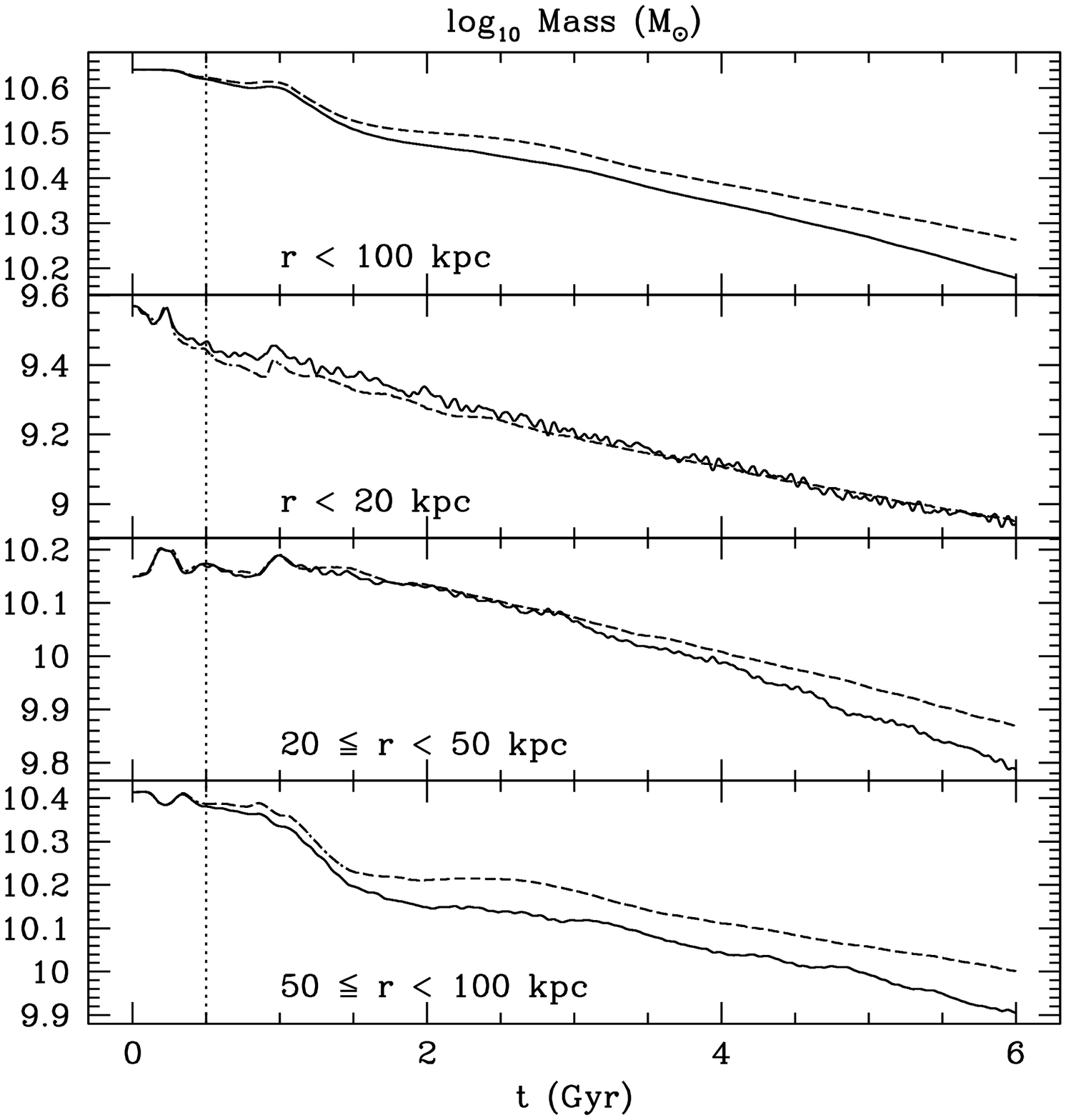}
\caption{Evolution of the intrinsic ISM mass for Run 5 with smaller turbulent driving 
scale ({\it dashed line}) in Case A, compared to the original run ({\it solid line}). 
The dotted line corresponds to 0.5 Gyr when the inflow of the ICM starts to enter the simulation box.
}
\label{fig:mass_ISM_add}
\end{figure}

\begin{figure}
\includegraphics[scale=0.43]{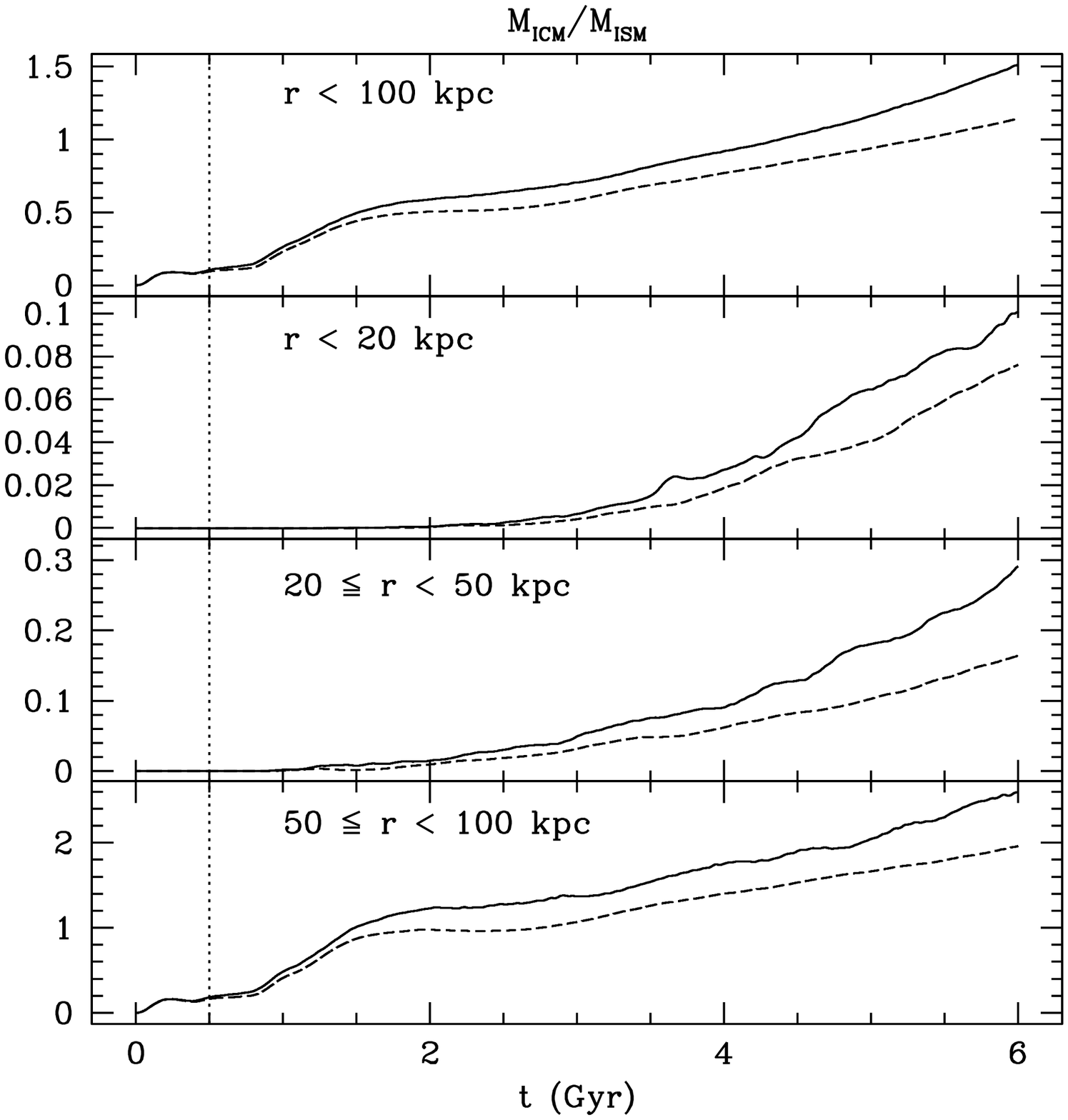}
\caption{Evolution of the ratio of the intrinsic ICM mass 
to the intrinsic ISM mass inside ${R_{t}}$ in Case A 
for Run 5 with smaller turbulence driving scale ({\it dashed line}), 
compared to the original run ({\it solid line}).
The dotted line corresponds to 0.5 Gyr when the inflow of the ICM starts to enter the simulation box.
}
\label{fig:mass_ICM_over_ISM_add}
\end{figure}

As shown in Figures \ref{fig:mass_ISM_add} and \ref{fig:mass_ICM_over_ISM_add}, 
the new simulation shows less stripping and weaker mixing between the ICM and ISM 
than in the original Run 5. 
The mass-weighted root-mean-square 1D Mach number of the ISM in the new simulation 
is about 0.1 which is almost the same as in the original Run 5. This Mach number 
corresponds to the state of the ISM before the onset of the ICM inflow. Since the 
energy injection rate is unchanged, and thermal energy dominates over the time-integrated 
dissipation rate of the injected energy, the Mach number does not change significantly.
Since the new simulation lacks large-scale motions in the ISM, the 
diffusion of the ISM with the ICM becomes inefficient. 
At 6 Gyr, about 20\% larger amount of the ISM survives 
stripping in the new simulation than in the original Run 5. 
Figure \ref{fig:mass_ICM_over_ISM_add} shows that 
the suppression of gas stripping is relatively strong at 
intermediate radii. Specifically,  for 20 $\le$ r $<$ 50 kpc, two times less 
ISM is retained in this shell at 6 Gyr.

\begin{figure*}
\includegraphics{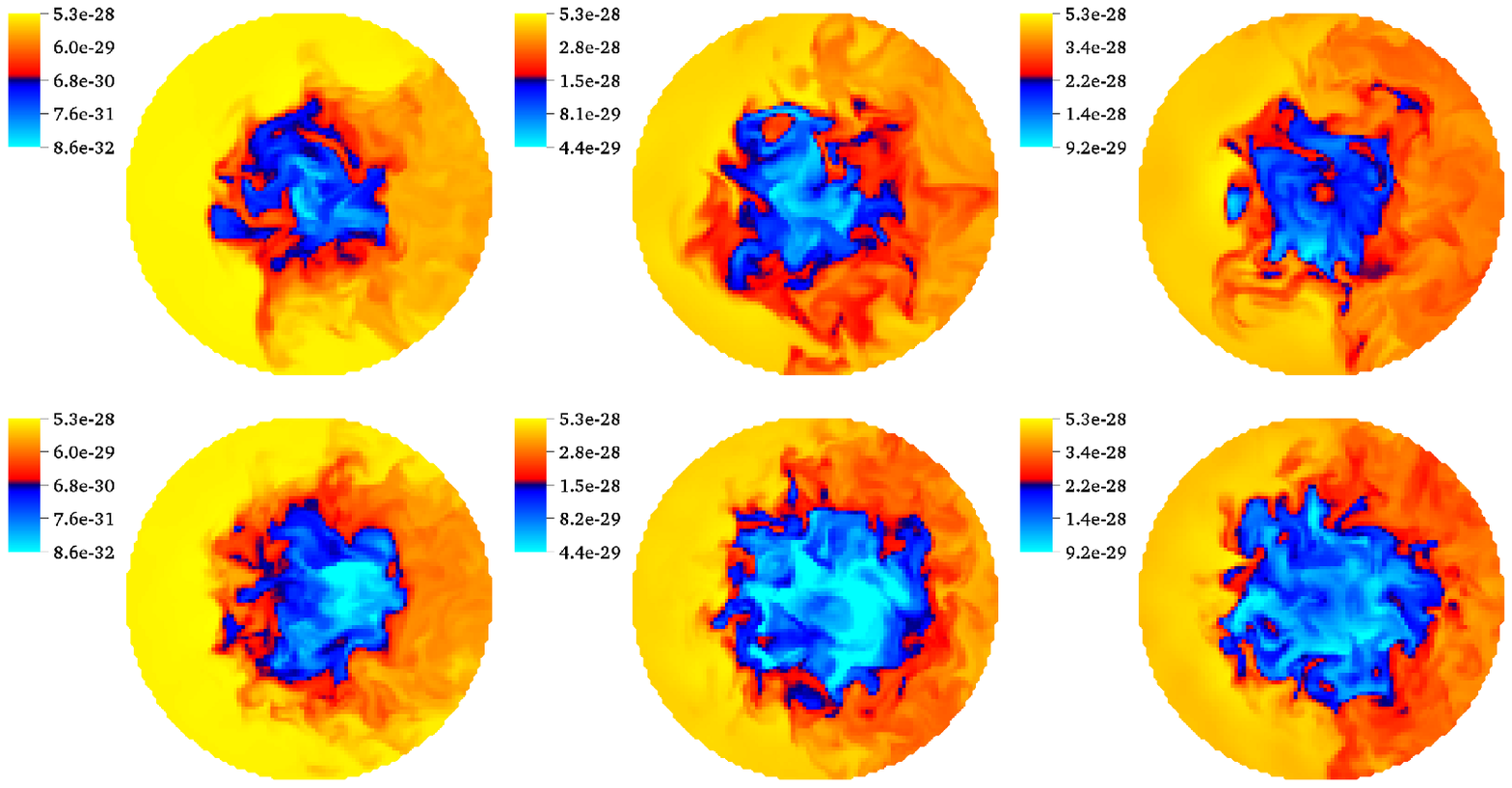}
\caption{ICM density distributions on the x-y plane centred on the galactic centre inside $R_{t}$, 
and parallel to the direction of the ICM inflow. The ICM flows from left to right. 
Columns correspond to 2, 4.5, and 6 Gyr ({\it from left to right}), respectively. 
Top row corresponds to the original Run 5 for Case A. Bottom row shows the result for the new 
simulation where the turbulence driving scale is about two times smaller than the original Run 5. 
The unit of density is ${\rm g ~ cm^{-3}}$. 
Note that the maximum in the colour maps is fixed in all panels, while 
the minimum varies from left to right. The minimum density is the same for a given time 
to allow for straightforward comparison of the effect of the turbulence driving scale on mixing.
}
\label{fig:2D_ICM_density}
\end{figure*}

Figure \ref{fig:2D_ICM_density} shows the ICM density distribution in a plane centred 
on the galaxy and parallel to the direction of the ICM inflow. Comparison of the 
panels in the same columns demonstrates that mixing is reduced in the new simulation: 
the area occupied by lower density gas is larger, and the penetration of the ICM 
deeper into the galactic potential is suppressed.

\end{document}